\documentclass{PoS}
\usepackage{tikz}
\usepackage{pgfplots}
\usetikzlibrary{shadows, calc}
\usetikzlibrary{arrows,shapes,positioning}
\usetikzlibrary{decorations.markings}
\usetikzlibrary{decorations.pathmorphing}
\usetikzlibrary{decorations.pathreplacing}
\tikzstyle arrowstyle=[scale=1]
\tikzstyle directed=[postaction={decorate,decoration={markings, mark=at position .65 with {\arrow[arrowstyle]{stealth}}}}]
\tikzstyle end directed=[postaction={decorate,decoration={markings, mark=at position 1 with {\arrow[arrowstyle]{stealth}}}}]
\tikzstyle reverse directed=[postaction={decorate,decoration={markings, mark=at position .65 with {\arrowreversed[arrowstyle]{stealth};}}}]
\tikzstyle{ann} = [fill=white,font=\footnotesize,inner sep=1pt] 

\tikzset{
  particle/.style={thick, postaction={decorate},
    decoration={markings, mark=at position 0.6 with {\arrow{triangle 45}}}},
  antiparticle/.style={thick, postaction={decorate},
    decoration={markings, mark=at position 0.4 with {\arrowreversed{triangle 45}}}},
  photon/.style={decorate, decoration={coil, aspect=0}},
  gluon/.style={decorate, decoration={coil, segment length=5pt, amplitude=4pt}},
  scalar/.style={decorate, decoration={border, segment length=6pt,amplitude=3pt, angle=0}},
  partgen/.style={thick, ->}
}
\definecolor{darkgreen}{rgb}{0,0.5,0}

\title{Nucleon spin and quark content at the physical point}

\ShortTitle{Nucleon spin and quark content at the physical point}

\author{\speaker{Constantia Alexandrou}\\%
      Department of Physics, University of Cyprus, CY-1678 Nicosia, Cyprus and\\
Computation-based Science and Technology Research Center, The Cyprus Institute, 20, K. Kavafi Str., 2121  Nicosia, Cyprus\\
      E-mail: \email{alexand@ucy.ac.cy}}
\author{Martha Constantinou, Kyriakos Hadjiyiannakou, Christos Kallidonis, Giannis Koutsou\\
Computation-based Science and Technology Research Center, The Cyprus Institute, 20, K. Kavafi Str., 2121  Nicosia, Cyprus\\
E-mail: \email{marthac@temple.edu}, \email{k.hadjiyiannakou@cyi.ac.cy},\email{c.kallidonis@cyi.ac.cy},\email{g.koutsou@cyi.ac.cy}}
\author{Karl Jansen, Christian Wiese\\
NIC, DESY, Platanenallee 6, D-15738 Zeuthen, Germany\\
E-mail: \email{karl.jansen@desy.de}}
\author{Alejandro Vaquero Avil\'es-Casco\\
INFN Sezione di Milano-Bicocca, Edificio U2, Piazza della Scienza 3, 20126 Milano, Italy\\
E-mail: \email{Alejandro.Vaquero@mib.infn.it}}

\abstract{We present results on the spin and quark content of the nucleon using $N_f=2$ twisted mass clover-improved fermion simulations with a pion mass close to its physical value. We use  recently developed methods to obtain accurate results for both connected and disconnected contributions. We provide results for the axial charge, quark and gluon momentum fraction as well as the light, strange and charm $\sigma$-terms.}

\FullConference{34th annual International Symposium on Lattice Field Theory\\
		 24-30 July 2016\\
		 University of Southampton, UK}

\begin{document}

\section{Introduction}
The contribution of quarks and gluons to the spin of the nucleon has
being studied experimentally since a number of years. Measurements
indicate that quarks contribute only about half to the spin of the
nucleon. Although the quark contribution to the spin of the nucleon
can be computed using lattice QCD from first principles, it is only
very recently that such a calculation has become possible, without any
approximations and directly at the physical value of the pion mass. We
present our recent results on the intrinsic spin $\Delta \Sigma^q$ and
total spin $J^q$ taking into account disconnected contributions, which
at the physical point cannot be neglected.

Another important quantity is the quark contents of the nucleon given
by $\sigma_f \equiv m_f\langle N|{\bar q}_f q_f|N\rangle$, where $q_f$
is a quark field of flavor $f$ and $|N\rangle$ is the nucleon
state. Beyond probing chiral symmetry breaking, a precise evaluation
of these quantities will reduce the uncertainty in interpreting
experiments for dark matter searches since the Higgs-nucleon coupling
depends on $\sigma_f$~\cite{Ellis:2008hf}. We present here a direct
evaluation of the light quark $\sigma_{\pi N}$, the strange quark
$\sigma_s$ and charmed quark $\sigma_c$ contents of the nucleon. In
this direct evaluation one computes the three-point function of the
scalar operator at zero momentum transfer, which additionally yields
the scalar charge after appropriate renormalization. Since charges are
the simplest matrix elements to evaluate, we start by discussing our
results on the scalar and tensor charges, as well as the axial charge
which yields the contribution of intrinsic quark spins to the nucleon
spin. The results are based on an ensemble of $N_f=2$ twisted mass
fermions with clover improvement, on a lattice size of $48^3\times 96$,
lattice spacing $a$ = 0.093(1)~fm and pion mass $m_\pi$ = 131~MeV,
referred to hereon as the physical
ensemble~\cite{Abdel-Rehim:2015pwa,Abdel-Rehim:2015owa,Abdel-Rehim:2014nka}.

\section{ Nucleon charges: $g_A$, $g_s$, $g_T$}
We consider the nucleon matrix elements $\langle
N(\vec{p^\prime}){\cal O}_X N(\vec{p})\rangle|_{q^2=0}$ of the scalar
operator, ${\cal O}_{S}^a=\bar{\psi}(x)\frac{\tau^a}{2}\psi(x)$ the
axial-vector operator, ${\cal
  O}_{A}^a=\bar{\psi}(x)\gamma^{\mu}\gamma_5\frac{\tau^a}{2}\psi(x)$
and the tensor operator ${\cal
  O}_{T}^a=\bar{\psi}(x)\sigma^{\mu\nu}\frac{\tau^a}{2}\psi(x)$. The
isovector combination, given by taking the Pauli matrix $\tau^3$, has
no disconnected contributions and we therefore first discuss the
isovector charges. While the axial charge is well-known
experimentally, $g_T$ and $g_S$ are not, however there is an
experimental program at JLab to improve the accuracy on the value of
$g_T$.

\begin{figure}
\begin{minipage}{0.49\linewidth}
{\includegraphics[width=\linewidth]{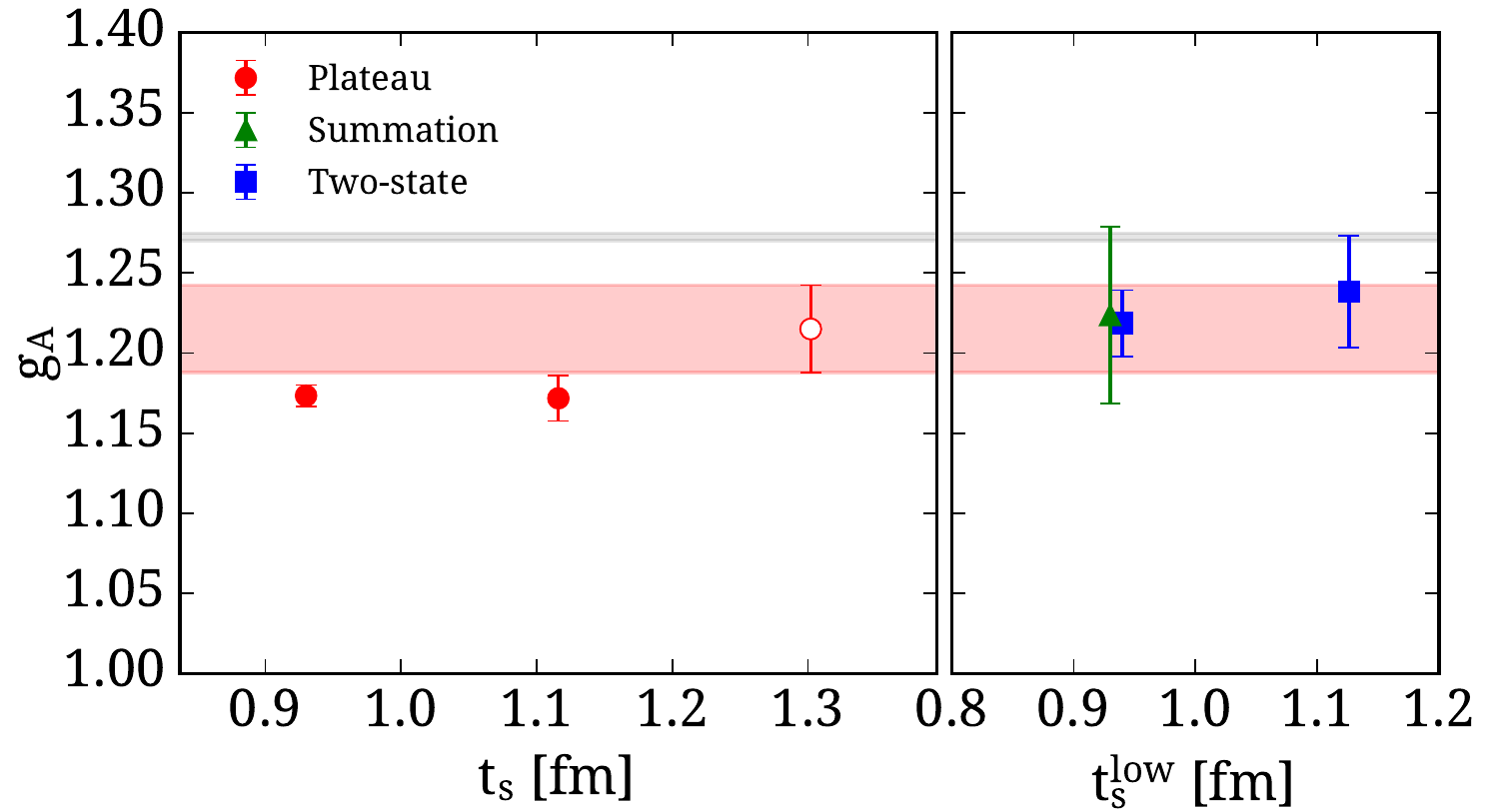}}
\end{minipage}
\begin{minipage}{0.46\linewidth}
\hspace*{-0.2cm} {\includegraphics[width=\linewidth]{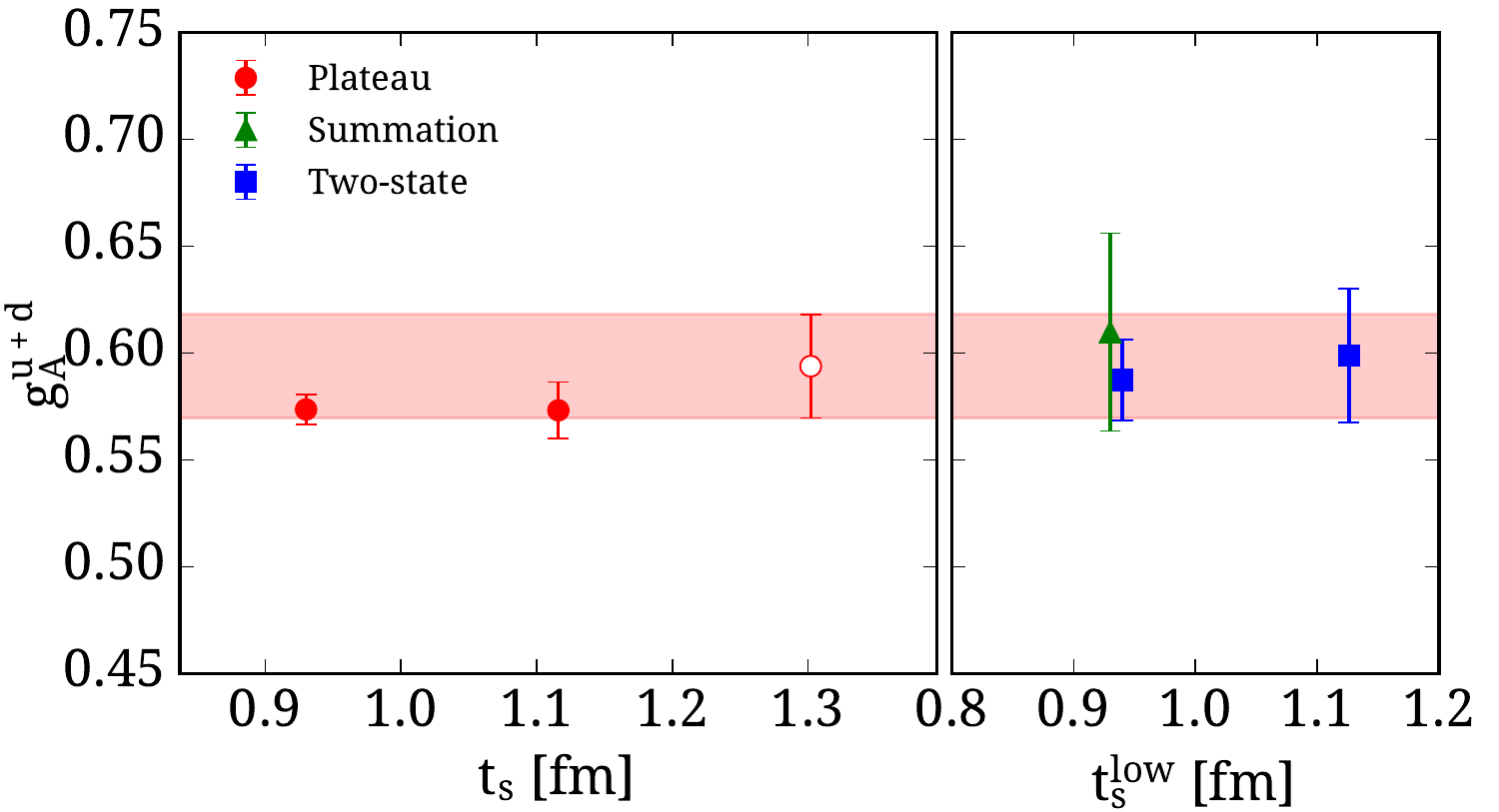}}
\end{minipage}
\caption{Isovector (left) and isoscalar connected (right) axial charge for three sink-source time separations: $t_s/a=10, 12, 14$, using 9260 measurements. $t_s^{\rm low}$ is the lowest value of $t_s$ used in the fits. The grey band is the experimental value. The open symbol is the value we take.}
\label{fig:axial}
\end{figure}
In Fig.~\ref{fig:axial} we show the value of the isovector and
isoscalar axial charge as we change the sink-source time separation
$t_s$ form $0.93$~fm to 1.3~fm, as well as, the results when using the
summation method and a fit including the first excited state
(two-state fit). For $t_S/a=14$ there is an increase in the value of
$g_A$, which however remains below the experimental value while the
two-state fit using $t_s/a=12$ and $t_s/a=14$ and the summation method
both having a larger error yielding consistent values. The connected
contribution to the isoscalar axial charge shows convergence, with
results from all methods being in agreement. We perform a similar
analysis for $g_S^{u-d}$ and $g_T^{u-d}$, where for the scalar we
increase $t_s$ to 1.7~fm in order to sufficiently suppress excited
states.  We used statistics of $\sim 9260$ for $t_s/a=10,$ $12,$ and
$14$, $\sim 48 000$ for $t_s/a=16$ and $\sim 70 000$ for $t_s/a=18$ to
keep the errors approximately constant. The values are listed in Table
I.

\begin{figure} 
 \begin{minipage}{0.33\linewidth}
\includegraphics[width=\linewidth]{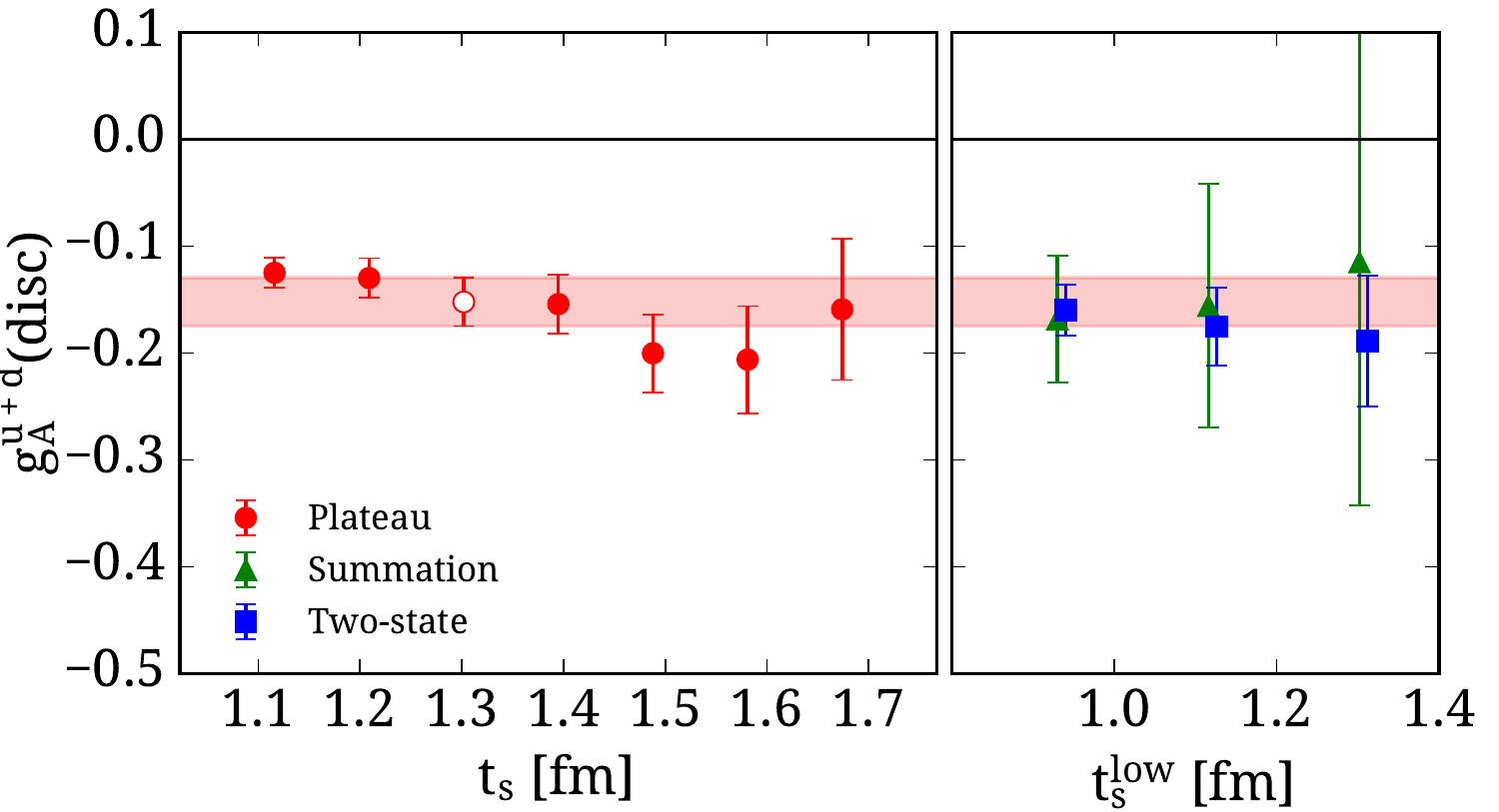}
\end{minipage}
\begin{minipage}{0.33\linewidth}
\includegraphics[width=\linewidth]{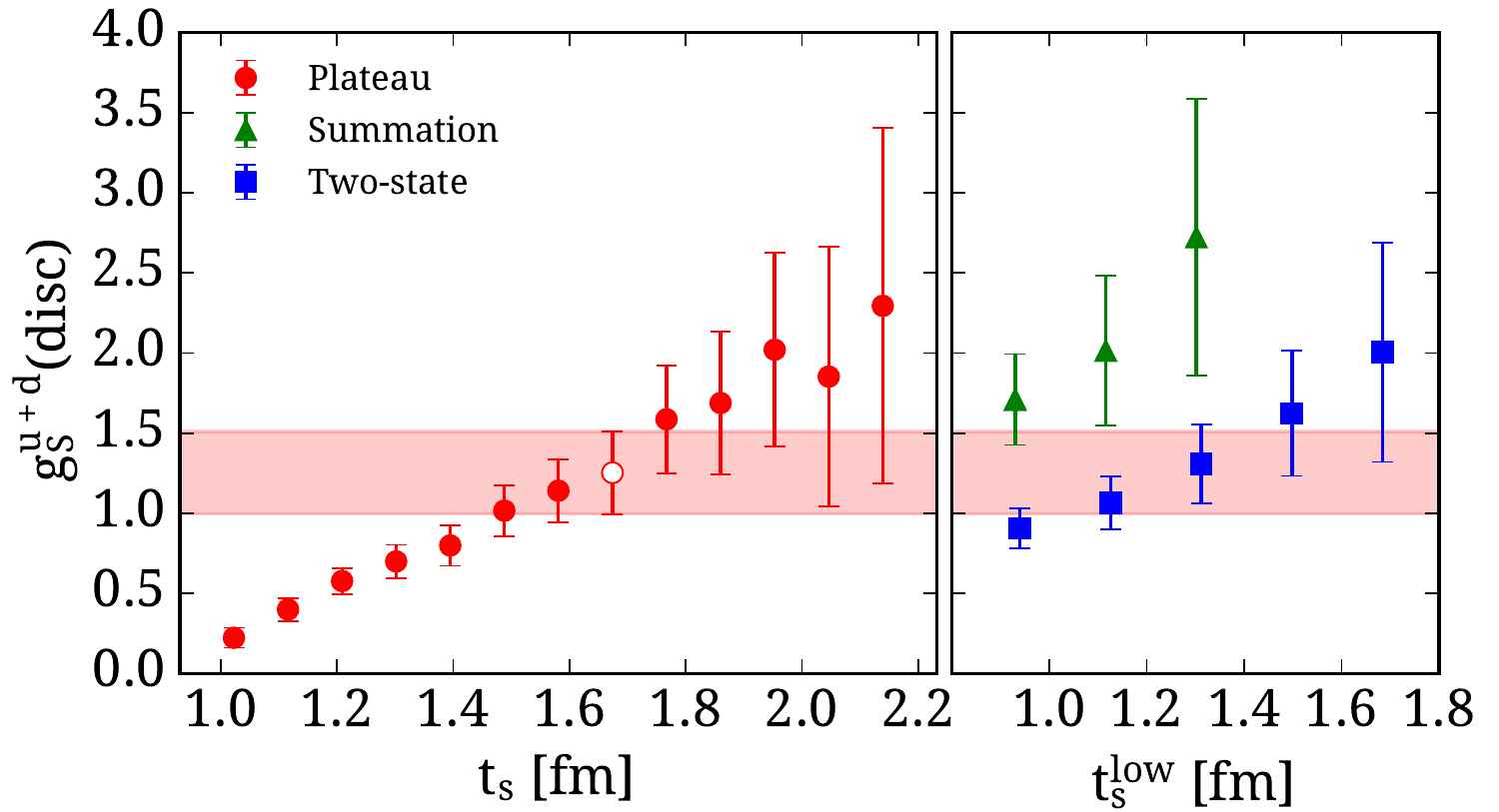}
   \end{minipage}\hfill
 \begin{minipage}{0.33\linewidth}
   \includegraphics[width=\linewidth]{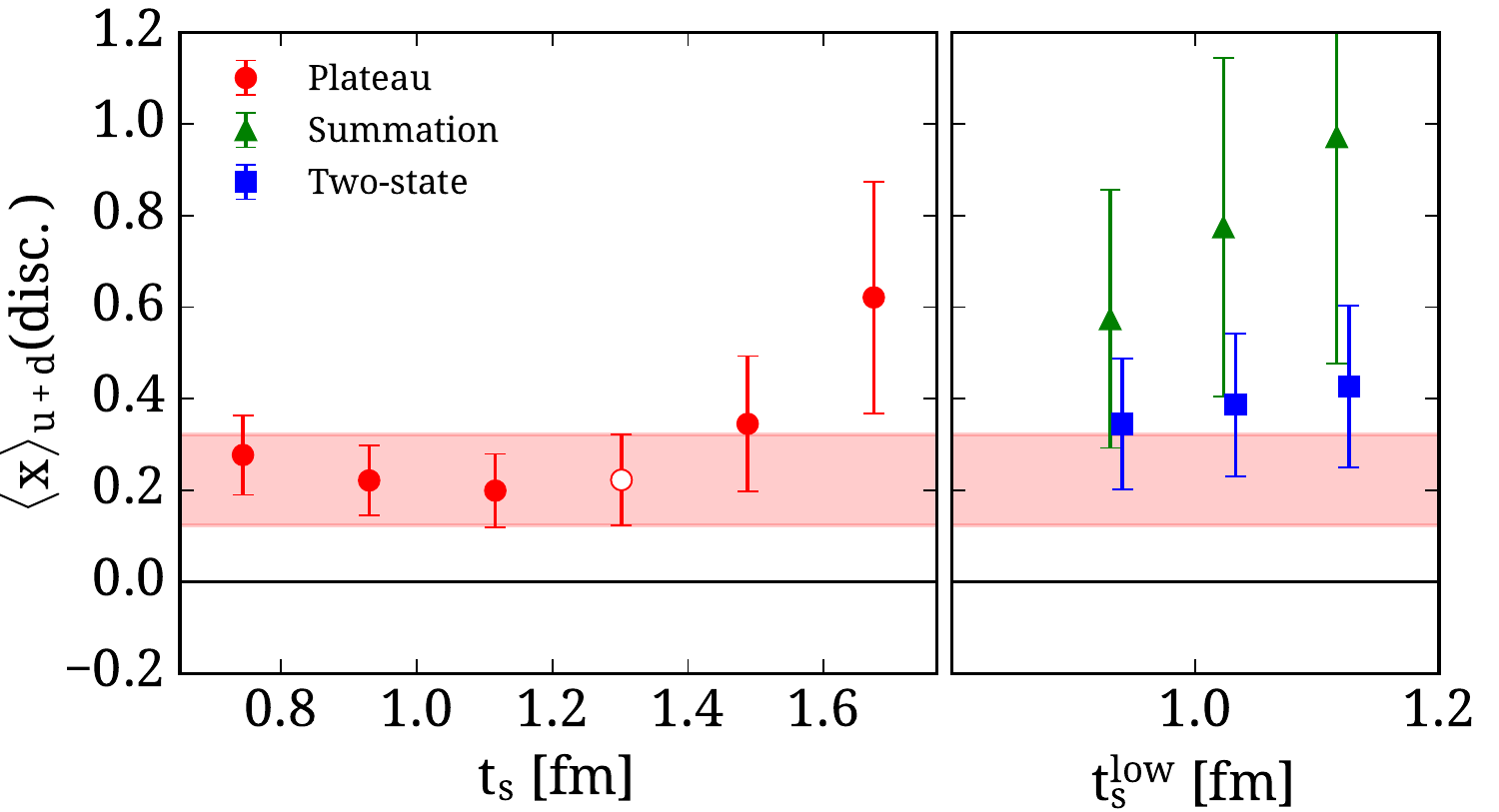}
  \end{minipage}
\caption{Disconnected contributions to the isoscalar axial charge
  (left), to the scalar charge (middle) and to the momentum fraction
  (right) shown as a function of $t_s$.}
\label{fig:A S}
\end{figure}
Besides the connected contributions, isoscalar nucleon charges also
receive disconnected contributions. Combining deflation of the
low-lying eigenvalues and the truncated solver method (TSM) we are
able to compute these quantities to sufficient accuracy for the first
time at the physical point~\cite{Vaquero}. GPUs are used to increase
statistics with 100 source positions computed per gauge configuration
for which we have checked that the error preserves the proper scaling.
In Fig.~\ref{fig:A S} we show the disconnected contributions to
$g_A^{u+d}$ and $g_S^{u+d}$ as we vary $t_s$ as well as the values
from the summation method and the two-state fit. As expected, for the
scalar one needs to increase $t_s$ in order to suppress excited states
sufficiently to isolate the nucleon matrix element. In both the scalar
and axial-vector the disconnected contribution are non-zero and need
to be taken into account. Disconnected contributions to the tensor
charge are consistent with zero. The statistics used for the
evaluation of these disconnected contributions are approximately
850,000.



\begin{table}[h]
\begin{center}
\small
\begin{tabular}{ccccc}
\hline
 $g_A^{u-d}$ & $g_A^{u+d}$ (conn.) & $g_A^{u+d}$ (disconn.)& $g_A^s$ & $g_A^c$ \\
1.21(3)(3) & 0.591(24) & -0.15(2) & -0.042(10) & -0.0053(30) \\
\hline
$g_S^{u-d}$ & $g_S^{u+d}$ (conn.) & $g_S^{u+d}$ (disconn.) &$g_S^s$ & $g_S^{c}$  \\
 0.93(25)(19) & 8.246(510) & 1.253(258) & 0.330(69) & 0.054(13)\\
\hline
 $g_T^{u-d}$ & $g_T^{u+d}$ (conn.) & $g_T^{u+d}$ (disconn.) & $g_T^s$ & \\
 1.004(21)(20) & 0.585(16) &  0.0007(11) & -0.0004(6)& \\
\hline
 $\langle x\rangle_{u-d}$ &  $\langle x\rangle_{u+d}$(conn.) &$\langle x\rangle_{u+d+s}$ & $\langle x\rangle_s$ & \\
 0.194(9)(10) & 0.498(13)  & 0.74(10) & 0.092(41)&\\
\hline
$\langle x\rangle_{\Delta u-\Delta d}$ &  $\langle x\rangle_{\Delta u+\Delta d}$ &  $\langle x\rangle_{\Delta u+\Delta d}$ (disconn.) & $\langle x\rangle_{\Delta s}$&\\
  0.259(9)(10) & 0.175(8) &-0.011(24)&  -0.051(49)&   \\
\hline
$\langle x\rangle_{\delta u-\delta d}$ & $\langle x\rangle_{\delta u+\delta d}$ (conn.) & $\langle x\rangle_{\delta u+\delta d}$ (disconn.)&  &  \\
0.273(17)(18) & 0.232(20) & -0.0038(51) &\\
\hline
\end{tabular}
\caption{\normalsize Values of the nucleon charges and first moments
  in the ${\overline{\rm MS}}$ scheme at 2~GeV for the physical point
  ensemble. For the isovector quantities the first error is
  statistical and the second systematic determined by the difference
  between the values from the plateau and two-state fits. }
\label{Table:values}
\end{center}
\end{table} 

\normalsize

\begin{figure}
\begin{minipage}{0.48\linewidth}
 \includegraphics[width=\linewidth]{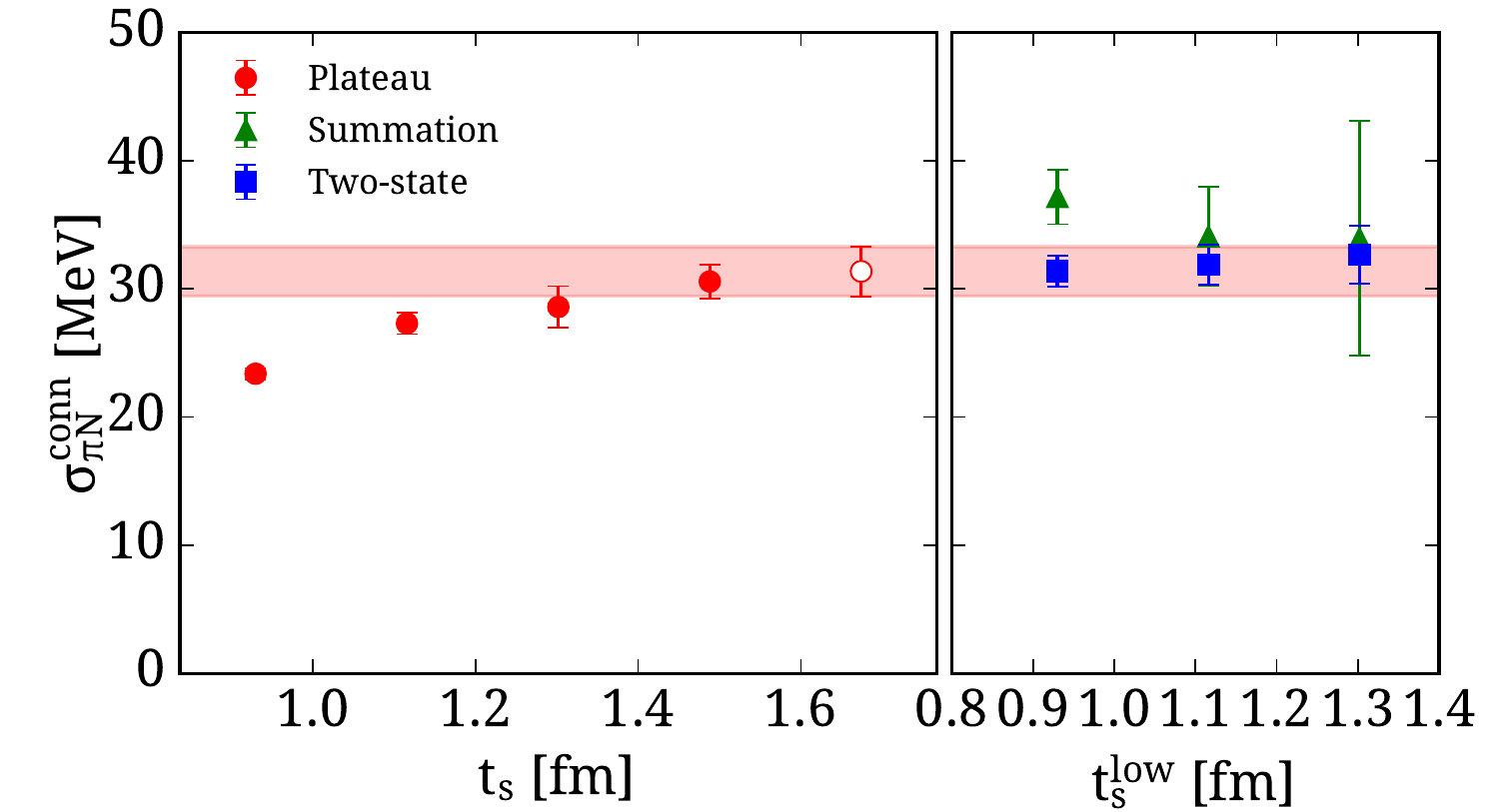}
\end{minipage}
\begin{minipage}{0.48\linewidth}
\includegraphics[width=\linewidth]{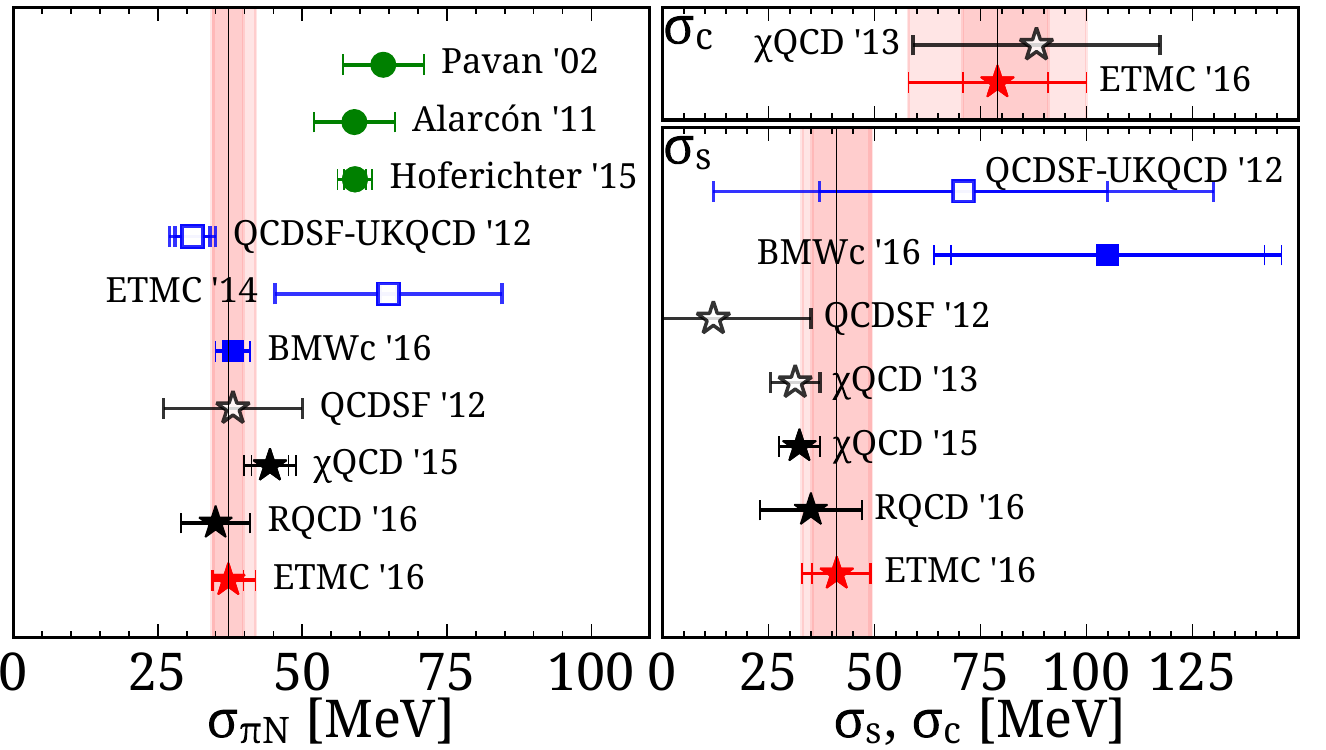}
\end{minipage}
\caption{On the left we show the convergence of the connected contribution to $\sigma_{\pi N}$. On the right we show  recent lattice QCD results on the light, strange and charm $\sigma$-terms.}
\label{fig:sigma}
\vspace*{-0.3cm}
\end{figure}

Nucleon $\sigma$-terms are computed either via the Feynman-Hellmann
theorem using $\sigma_q=m_l\frac{\partial m_N}{\partial m_q}$ or by a
direct computation of the scalar matrix element. We have computed the
light, strange and charm $\sigma$-terms using our physical
ensemble~\cite{Abdel-Rehim:2016won} by computing the nucleon matrix
element of the scalar operator as shown in Fig.~\ref{fig:A S} for the
connected contribution and in Fig.~\ref{fig:sigma} for the disconnected
contribution to $\sigma_{\pi N}$ and for $\sigma_{s}$ and $\sigma_{c}$
which are completely disconnected. With our increased statistics we
find $\sigma_{\pi N}= 36(2)~$MeV, $\sigma_s= 37(8)$~MeV and
$\sigma_c=83(17)$~MeV. Comparison with other recent lattice QCD
results and phenomenology is also shown in Fig.~\ref{fig:sigma}.

\vspace*{-0.3cm}

\section{First moments: $\langle x \rangle_q$, $\langle x \rangle_{\Delta q}$, $\langle x \rangle_{\delta q}$}

\vspace*{-0.3cm}

We consider the one derivative operators yielding the unpolarized moment
$\langle x \rangle_q=\int_0^1 dx \> x\left[q(x)+\bar{q}(x)\right]$,
the helicity moment,
$\langle x\rangle_{\Delta q}=\int_0^1 dx \> x\left[\Delta q(x)-\Delta \bar{q}(x)\right]$
and the transversity moment
$\langle x\rangle_{\delta q}=\int_0^1dx x\left[\delta q(x)+\delta \bar{q}(x)\right]$,
where  $q(x)=q(x)_\downarrow+q(x)_\uparrow$,  $\Delta q(x)=q(x)_\downarrow-q(x)_\uparrow$ and $\delta q(x)=q(x)_\perp+q(x)_\top$.

\begin{figure}
\begin{minipage}{0.49\linewidth}
 \includegraphics[width=\linewidth]{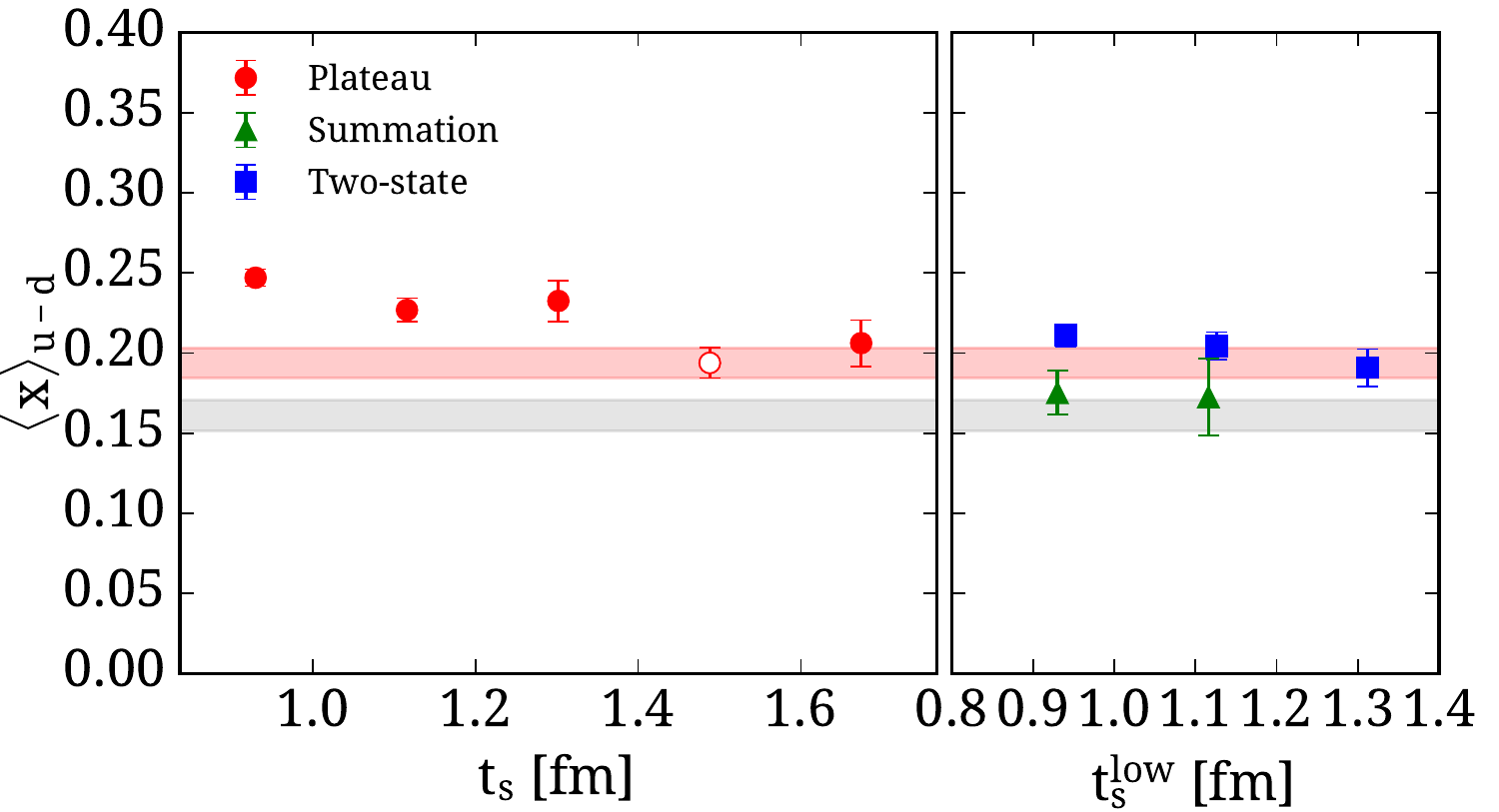}
\end{minipage}
\begin{minipage}{0.49\linewidth}
 \includegraphics[width=\linewidth]{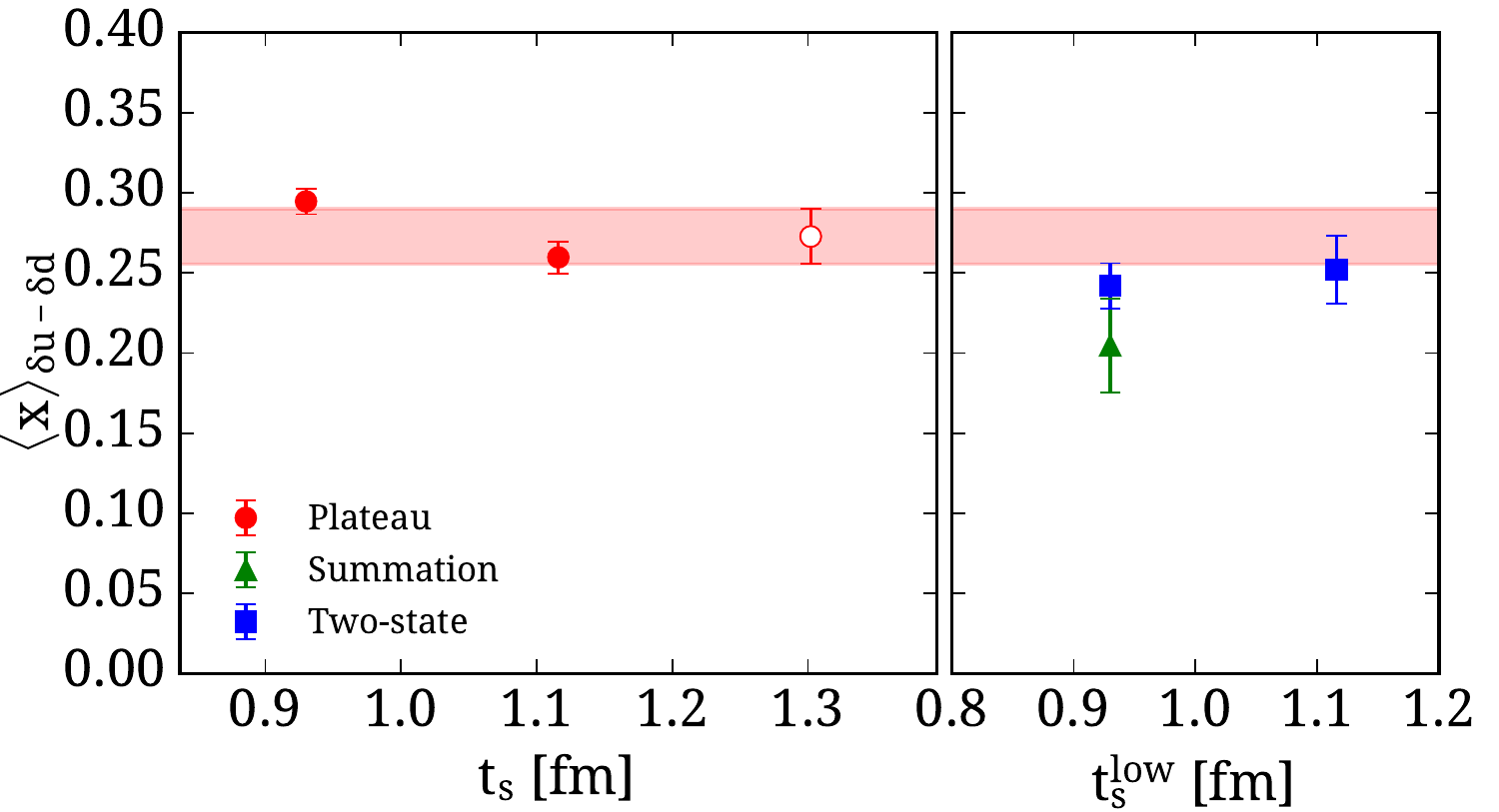}
\end{minipage}
\caption{The isovector unpolarized (left) and helicity moments (right). The notation is the same as that of Fig.~1.}
\label{fig:x isov}
\end{figure}

In Fig.~\ref{fig:x isov} we show the isovector unpolarized and
transversity moments as the sink-source time separation is increased,
as well as the results from the summation and the two-state fits. The
quark momentum fraction given by $\langle x\rangle_{u-d}$ decreases as
$t_s$ increases, as expected from previous studies, approaching the
experimental value. Similar behaviour is also observed for the
polarized moment $\langle x\rangle_{\Delta u - \Delta d}$. The
transversity for all three values of $t_s$ yields consistent results
that also agree with the values extracted using two-state fits,
pointing to less contamination due to excited states. Larger values of
$t_s$ are under study.
We have also computed the isoscalar quantities together with the disconnected contributions. These are given in Table I.  For the isoscalar $\langle x\rangle_{u+d+s}$ we take into account perturbatively the mixing with the gluon momentum fraction $\langle x\rangle_{g}$. Neglecting the mixing we find  $\langle x\rangle_{u+d+s}=0.81(11)$ and $\langle x\rangle_{u+d}=0.72(10)$.

\vspace*{-0.3cm}

\section{\bf Gluon content of the nucleon}
\vspace*{-0.3cm}

Gluons can carry a significant amount of momentum and spin in the
nucleon and thus far only a few studies have computed this
contribution~\cite{Yang:2016plb,Alexandrou:2013tfa}. In order to evaluate the gluon contribution to
the nucleon spin we need to compute $J_g = \frac{1}{2} (A_{20}^g +
B_{20}^g)$, that involves the momentum fraction $A_{20}^g = \langle x
\rangle_g$ and $B_{20}^g$. As a first step we compute here the gluon
momentum fraction from the nucleon matrix element of the gluon operator
$O_{\mu\nu}=-G_{\mu\rho} G_{\nu\rho}$. The gluon momentum fraction is
then extracted from $\langle N(0)|
O_{44}-\frac{1}{3}O_{jj}|N(0)\rangle=m_N \langle x \rangle_g$.

The disconnected correlation function connecting a gluon loop is known
to be very noisy. We thus employ several steps of stout smearing in
order to remove the UV fluctuations in the gauge field. The
methodology was tested for $N_f=2+1+1$ twisted mass fermions at
$m_\pi=373$~MeV~\cite{Alexandrou:2013tfa}.

\begin{figure}
\includegraphics[scale=0.8]{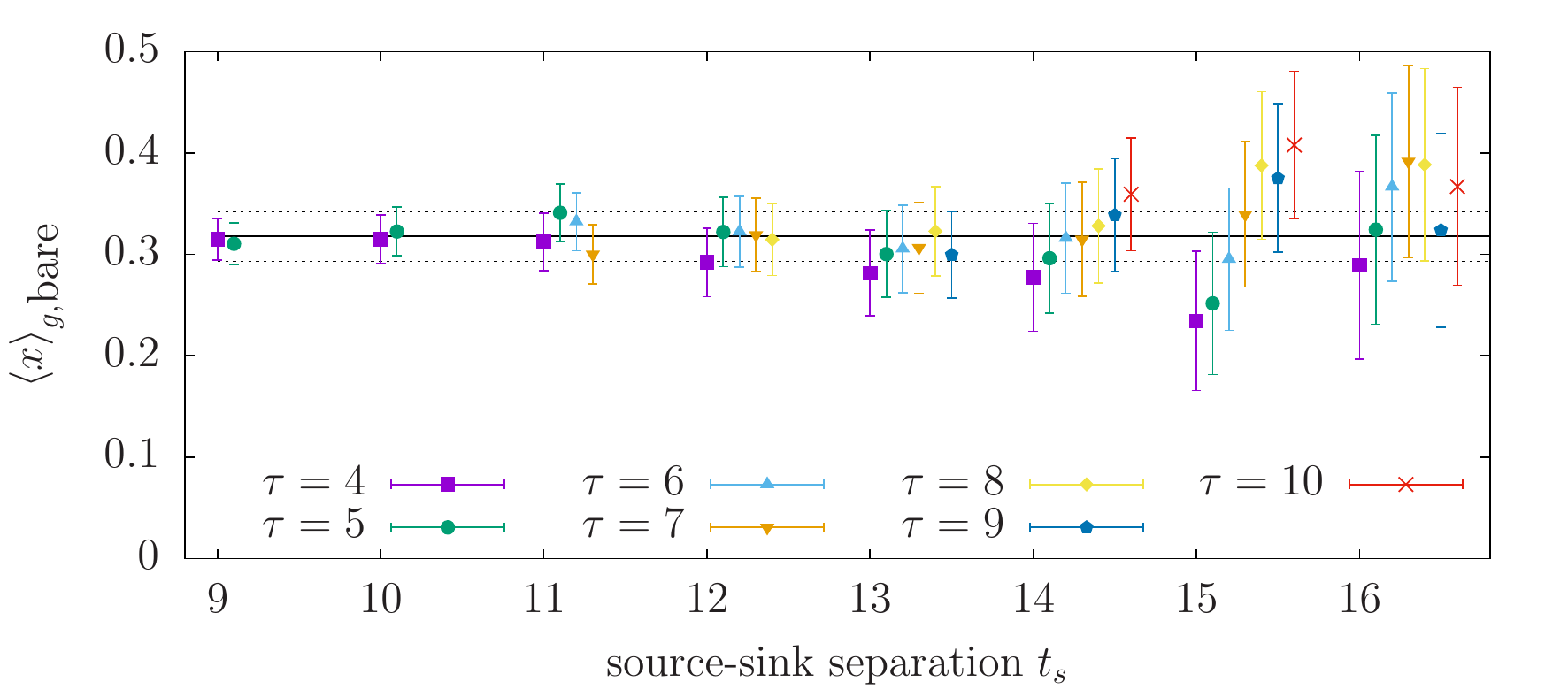}
\caption{ Gluon moment as a function of the sink-source time separation $t_s$ in units of the lattice spacing for various fits ranges $\tau$. }
\label{fig:gluon}
\end{figure}

Using the physical ensemble and analyzing 2094 gauge configurations
with 100 different source positions each, leading to over 200\,000
measurements, we obtain an accurate value for the bare gluon momentum
fraction as shown in Fig.~\ref{fig:gluon}, namely $\langle x
\rangle_{g,\mbox{bare}} = 0.318(24)$. Due to mixing with the quark
singlet operator, the renormalization and mixing coefficients have to
be extracted from a one-loop perturbative lattice
calculation. This yields $\langle x\rangle^R_g = Z_{gg}
\langle x\rangle_g + Z_{gq} \langle x\rangle_{u+d+s} =
0.273(23)(24)$. The renormalization is carried out perturbatively using
two-levels of stout smearing. The systematic error is the difference
between using one- and two-levels of stout smearing. Having both the
quark and gluon momentum fractions we can check whether the momentum
sum is satisfied. We find: $\sum_q\langle x \rangle_q+\langle x
\rangle_g=\langle x \rangle_{u+d}^{\rm conn}+\langle x
\rangle_{u+d+s}^{\rm disconn}+\langle x \rangle_g=1.01(10)(2)$, which indeed
is consistent with unity.

\section{Nucleon spin}
An analysis of the proton spin decomposition has been an important
long-term goal for understanding the origin of spin in the nucleon,
particularly after the the European Muon
Collaboration~\cite{Ashman:1989ig} showed that the quark spin only
contributes a surprisingly small amount to the total spin of the
proton. Using our results on $g_A^q=\Delta \Sigma^q$ computed
including disconnected contributions at the physical point we can
compare directly to the experimental results.

\begin{figure}[h]
\begin{minipage}{0.49\linewidth}
{\includegraphics[width=0.98\linewidth]{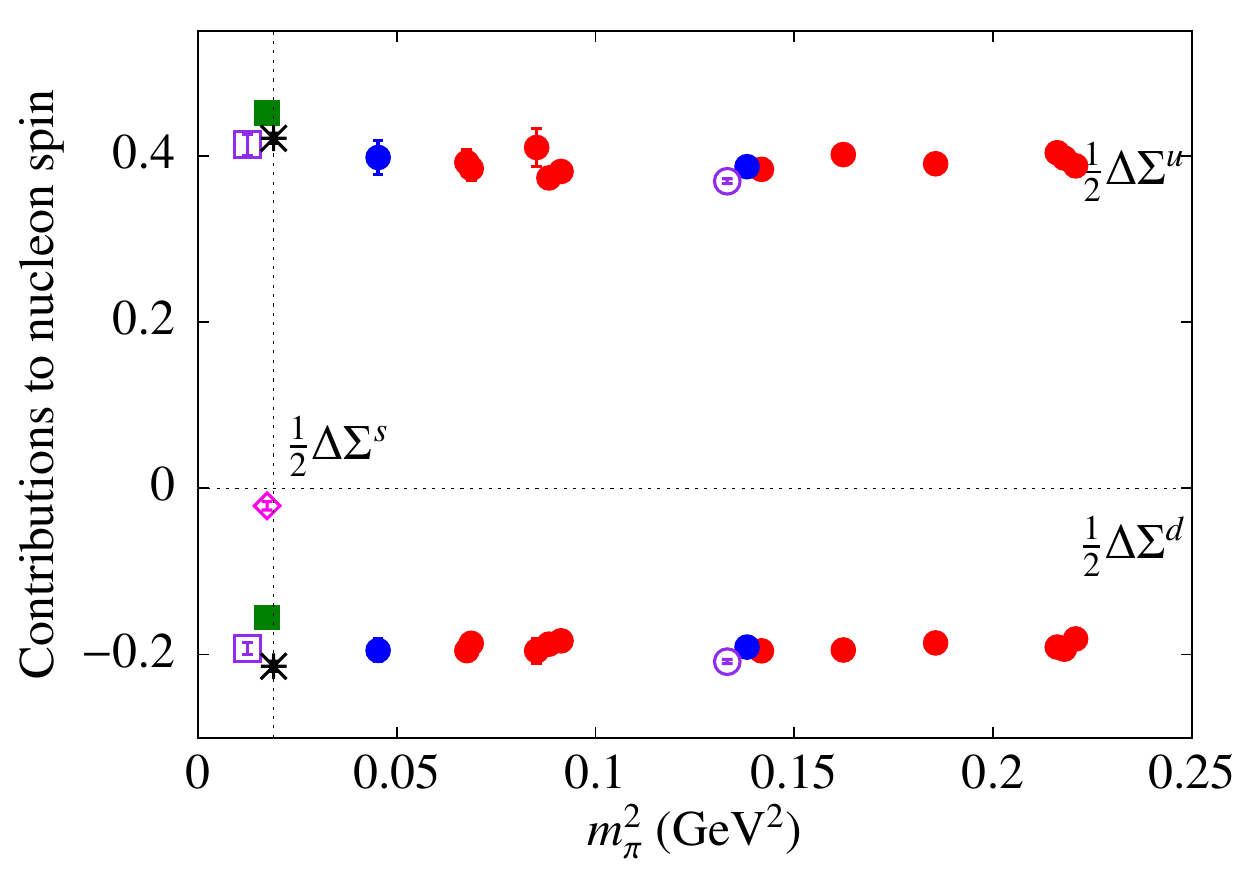}}
\caption{$\Delta \Sigma^q$ as a function of the pion mass using
 twisted mass fermions (TMF). }
\label{fig:spin} 
\end{minipage}\hfill
\begin{minipage}{0.47\linewidth}
In Fig.~\ref{fig:spin} we show $\frac{1}{2}\Delta\Sigma^q$ for various
$N_f=2$ (red circles) and $N_f=2+1+1$ (blue circles) TMF ensembles. The value for the physical ensemble 
 is shown with the green squares and the experimental value with the asterisks.
 For the physical ensemble as
well as for an $N_f=2+1+1$ ensemble with $m_\pi=373$~MeV (B55) we include
the effect of the disconnected contribution, which although small, shifts
the values towards the experimental one.
\end{minipage}
\end{figure}

\begin{figure}
\begin{minipage}{0.49\linewidth}
{\includegraphics[width=\linewidth]{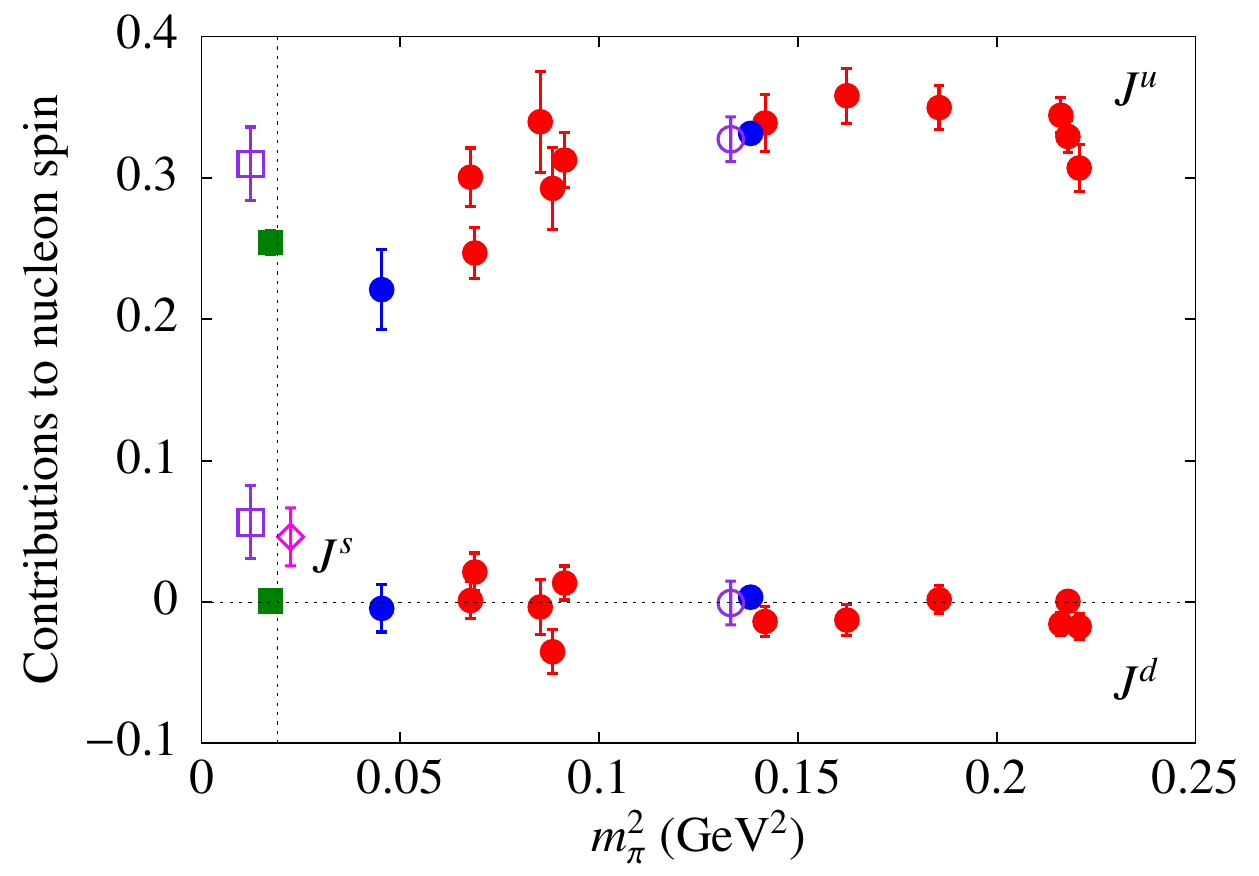}}
\end{minipage}
\begin{minipage}{0.49\linewidth}
{\includegraphics[width=\linewidth]{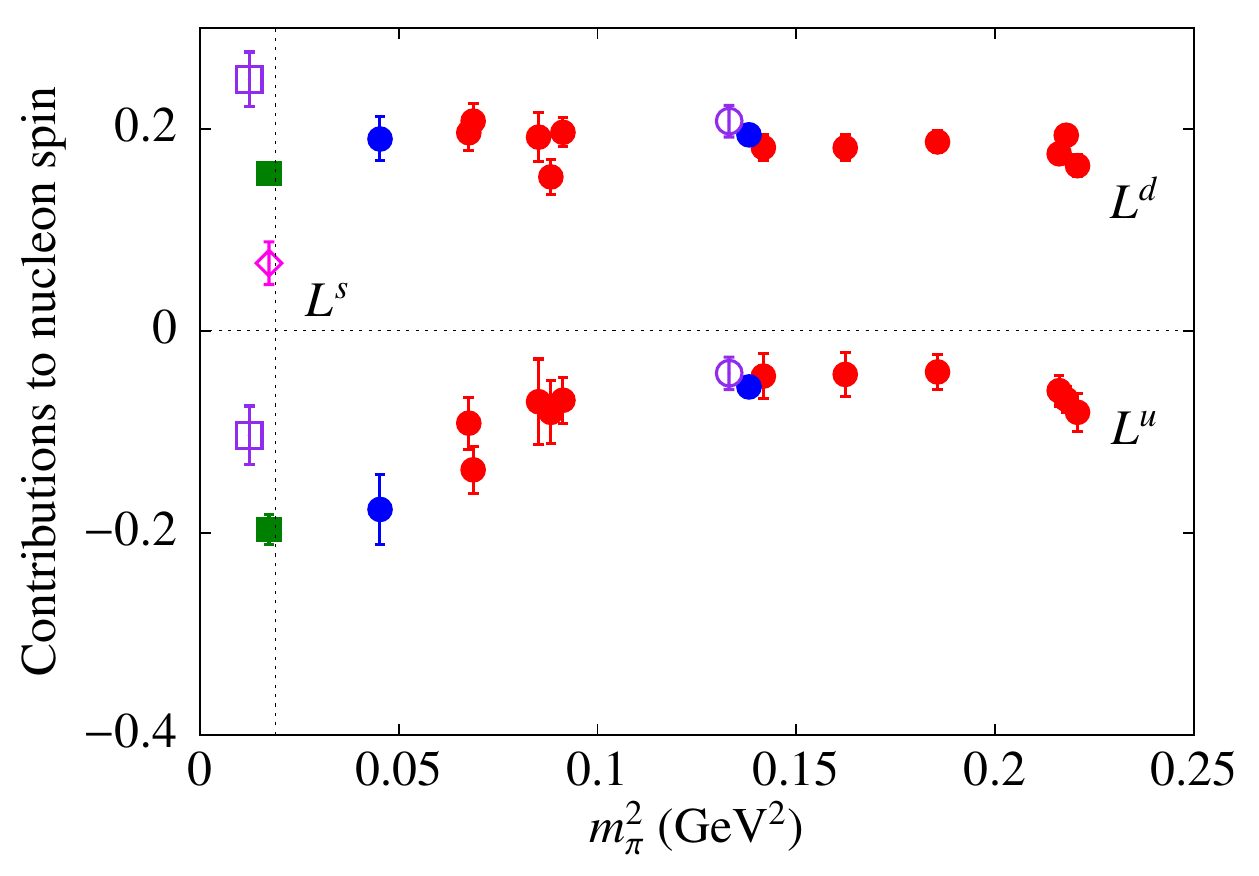}}
\end{minipage}
\caption{The total quark spin (left) and angular momentum $L^q$ (right) as a function of $m_\pi$. The notation is the same as in Fig.~6.}
\label{fig:J L}
\vspace*{-0.3cm}
\end{figure}

The total quark and gluon spin
$J^{q,g}=\frac{1}{2}\left(A_{20}^{q,g}(0)+B_{20}^{q,g}(0)\right)$ can
be extracted from the nucleon matrix element of the one derivative
vector operator that yields the generalized form factors $A_{20}(0)$
and $B_{20}(0)$. While $A_{20}$ can be extract directly at zero
momentum transfer, $B_{20}$ must be extrapolated using results at
finite momentum transfers. We found that disconnected contributions to
$B_{20}(0)$ are consistent with zero and we therefore do not include
them in this analysis. Disconnected contribution to $A_{20}$ are
non-zero and are included for the physical and B55 ensembles.  Using
the spin sum rule $ J^q=\left(\frac{1}{2}\Delta \Sigma^q +L^q\right) $
one can extract the orbital angular momentum $L^q$. The spin $J^q$
and $L^q$ are shown in Fig.~\ref{fig:J L}~\cite{Alexandrou:2016mni}.
We note that the isovector and isoscalar momentum fraction are
normalized with the same non-perturbatively determined renormalization
constant neglecting the small mixing with the gluon momentum so that
we can extract the individual quark contributions.

At the physical point we obtain:
\begin{eqnarray}
\frac{1}{2}\Delta \Sigma^u =0.413(13),\,\, & \frac{1}{2}\Delta \Sigma^d =-0.193(7),\,\, & \frac{1}{2}\Delta \Sigma^s =-0.021(5)\,\,\\ \nonumber
J^u =0.310(26),\,\, & J^d =0.056(26),\,\, & J^s =0.046(21)\\ \nonumber
L^u =-0.104(29),\,\, &  L^d =0.249(27),\,\, & L^s =0.067(21)
\end{eqnarray}
Using the spin sum rule $ \frac{1}{2}=\sum_{q=u,d,s}\left(J^q\right) +J^g $ we deduce that $J^g=0.09(6)$, which is in agreement with  the direct evaluation of the momentum fraction $\frac{1}{2}\langle x\rangle_g=0.136(12)$~\cite{Alexandrou:2016ekb}. This indicates that $B_{20}^g$ is small, which is in line with what is found for the quark connected contribution that yields $B_{20}^{u+d}=0.012(20)$. Neglecting $B_{20}^g$, we find for the nucleon spin, $J_N=\sum_{q=u,d,s}(J^{q})+J^g=0.51(5)(4)$, which satisfies the spin sum of $1/2$. The systematic error is the difference between neglecting i.e. using non-perturnbative renormalization and including perturbatively the mixing with $\langle x\rangle_g$. 

\vspace*{-0.3cm}

\section{Conclusions}

Using high statistics we are able to compute both connected and
disconnected contributions to the nucleon charges and to the first
moment of parton distributions. Excited states contamination is
carefully examined for all quantities. For the scalar charge and
correspondingly the $\sigma$-terms the analysis is carried out for
sink-source time separations up tot 1.7~fm to ensure that the nucleon
ground state dominates. For the axial charge, polarized and helicity
moments larger sink-source time separations are under investigation.
The quark spin decomposition shows that the intrinsic spin is in
agreement with the experimental values, while we obtain non-zero
disconnected contributions to the orbital angular momentum $L^q$.

\noindent
{\bf Acknowledgments:}
We used computational resources on Piz Daint at the Swiss Supercomputing Center,  under project ID s625 and s540, from the
John von Neumann-Institute for Computing on the Juropa
system and the BlueGene/Q system Juqueen at
the research center in J\"ulich, from the Gauss Centre
for Supercomputing on HazelHen (HLRS) and SuperMUC
(LRZ), and on the  Cy-Tera machine,
The Cyprus Institute, funded by RPF,
NEAYPODOMH/STPATH/0308/31. This project is partly supported by the EU  Horizon 2020 research and innovation program  Marie Sklodowska-Curie under grant agreement No 642069.

\end{document}